# On the Application of Optimal Control Theory to Climate Engineering

Sergei  Soldatenko and Rafael Yusupov

*St. Petersburg Institute for Informatics and Automation of the Russian Academy of Sciences,*

*No. 39, 14-th Line, St. Petersburg, 199178, Russia*

soldatenko@iias.spb.su

## Abstract

Current and projected human-induced global warming is one of the major threats for natural and anthropogenic systems in the XXI century. Technologies aiming at manipulating the earth's climate system, which are known as geoengineering or climate engineering, are considered as a potential option to decelerate the global warming and stabilize the climate. Although climate engineering actually aims to control the climate, geoengineering operations are still predominantly considered as intentional large-scale interventions on the climate system rather than techniques for climate control within the scope of the conventional control theory. In this paper, the new method is considered that allows identifying the "perfect" scenario of climate engineering operations based on the optimal control theory. The application of this approach is demonstrated using zero-dimensional energy-balance climate model in which the global average surface temperature represents the system's state variable and the albedo of aerosols injected into the stratosphere is designated as the control variable. Pontryagin's maximum principle is applied to find the best possible control and the associated climate system's trajectory for a specific objective function (performance measure). Discussed illustrative results were analytically and numerically calculated using the four greenhouse gas concentration scenarios based on the Representative Concentration Pathways. Apparently, the use of the optimal control theory can provide valuable information for the development of optimal strategies for climate manipulation in order to counter global warming.

# 1 Introduction

The ongoing increase in global average surface temperature induced by human activity is one of the greatest challenges facing mankind in the XXI century (IPCC 2013). To counter global warming, a wide range of technologies and methods known as geoengineering or climate engineering, have been proposed. The detailed review and analysis of climate engineering can be found in numerous publications (e.g., Goodell 2011; Jones at al. 2016; Keith 2013; Launder and Thompson 2010; MacCracken 2009; Ming et al. 2014; Stilgoe 2015; Shepherd et al. 2009; Wigley 2006). This article is not intended to provide an overview of geoengineering methods and approaches. Instead, we only emphasize that among many possible technologies solar radiation management (SRM) by injection of aerosols or aerosol precursors into the stratosphere represents one of the most realistic and feasible solutions. The goal of SRM is to induce a negative radiative forcing of the climate system and, consequently, to partially compensate the positive radiative forcing caused by growing concentrations of greenhouse gases (GHGs) in the atmosphere (e.g., Budyko 1974; Crutzen 2006; McClellan et al. 2012; Pope et al. 2012; Rasch et al. 2008; Robock et al. 2009). To be able to suggest geoengineering to fight global warming, we obviously have to assume that the climate system can be controlled by humans to some extent. Nevertheless, climate engineering and, in particular, SRM is primarily considered outside the framework of the conventional control theory (e.g., Bubnicki 2005; Isidori 1999; Zabczyk 2007). Effects of SRM interventions are commonly evaluated by computer simulations which involve the development of particular scenarios for SRM operations (e.g. Jarvis and Leedal 2012; Kravitz et al. 2015 and the references therein), although developing hypothetical scenarios is a process with an unavoidable element of subjectivity. Thus, from the standpoint of the control theory, the objectives of geoengineering operations and associated constraints and limitations are formulated rather vaguely. Meanwhile, climate engineering is a purposeful process having the objective that can be mathematically formulated in terms of an extremal problem which can be solved using methods of the optimal control theory (Bellman 1957; Kirk 1970; Pontryagin et al. 1962; Sontag 1990). This approach known as geophysical cybernetics was introduced in the late 1970s as a new multidisciplinary research area for planning and implementation of human deliberate interventions on dynamical, physical and chemical processes in the earth's climate system on the basis of the ideas and methods from cybernetics (Yusupov 1979). Ultimately, this means that the climate system and its subsystems and processes can be viewed as self-regulating control systems, while specifically-trained operators, equipped with special tangible resources, act as controller. Some ideas of geophysical cybernetics have been discussed by Schellnhuber and Kropp (1998) and Peng at al. (2002) in the context of global environmental and socio-economic development management. An optimal control problem (OCP) for weather modification and climate manipulation, both probabilistically and deterministically, has been conceptually formulated in our previous publications (e.g., Gaskarov et al. 1998; Soldatenko and Yusupov 2015a; Soldatenko and Yusupov 2015b; Soldatenko and Yusupov 2016; Soldatenko 2017) taking into account that the earth's climate is a large-scale physical system with unique properties and numerous positive and negative feedback loops (e.g., Hansen et al. 1984; Soden and Held 2006; Roe 2009). Therefore, viewed from the perspective of cybernetics, the climate system affected by natural external and internal factors that force climate dynamics can be considered to be a closed-loop control system. The earth's climate as a feedback control system has been examined by Jarvis at al. (2008, 2009) and MacMartin et al. (2014) with respect to SRM. It was shown that the representation of climate system in the form of a closed-loop control system effectively reduces the uncertainties of SRM operations. However, in these papers the problem has not been mathematically considered as an OCP with the formulation of the objective function that must be minimized or maximized. In a number of papers, the optimal control theory has been applied to find the optimal strategy for greenhouse gas emissions abatement and to solve the socio-economic planner's problem (e.g., Brock 2013; Greiner et al. 2014; Maurer et al. 2013; Maurer et al. 2015; Moles et al. 2003; Tol 2002), but climate engineering has not been the focus of these studies.

This paper aims to present the new method for identifying the "perfect" scenario of climate engineering operations based on the optimal control theory. To illustrate the application of this approach we adopt a zero-dimensional energy-balance model in which the global average surface temperature represents the system's state variable and the albedo of artificially created aerosol cloud is considered to be the control variable. Pontryagin's maximum principle is applied to find the optimal control (albedo of the aerosol layer) and the associated climate system's trajectory for specific objective function (performance measure). To evaluate an anthropogenic radiative forcing the Representative Concentration Pathways scenarios have been used. Presentation of the material will be carried out without excessive mathematical formalism. However, some basic concepts of the optimal control theory are included for completeness in the Appendix.

## 2    The model of control system

As a starting point we consider a zero-dimensional energy balance model of the form (e.g., McGuffie and Henderson-Sellers 2005; Karper and Engler 2013):

$$C\frac{dT}{dt} = R^{\downarrow} - R^{\uparrow}, \tag{1}$$

where $T$ is the planet's average surface temperature; $C$ is the effective heat capacity of the atmosphere-ocean system, measured in W m$^{-2}$ K$^{-1}$; $R^{\downarrow}$ and $R^{\uparrow}$ are fluxes of the shortwave solar radiation and longwave outgoing radiation respectively, which are parameterized as follows (e.g., Karper and Engler 2013):

$$R^{\downarrow} = Q\left[1 - \alpha\left(T\right)\right], \quad R^{\uparrow} = \varepsilon\sigma T^4. \tag{2}$$

Here, $Q$ is a solar insolation on the top of the atmosphere defined as $Q = I_0/4$, where $I_0$=1368 W m$^{-2}$ is a solar constant; $\alpha = 0.3$ is a planetary albedo; $\varepsilon \approx 0.62$ is a factor called planetary emissivity; $\sigma = 5.67 \times 10^{-8}$ W m$^{-2}$ K$^{-4}$ is the Stefan-Boltzmann constant. Following Budyko (1974), it is reasonable to approximate the outgoing longwave flux by a linear regression:

$$R^{\uparrow} = A + BT \tag{3}$$

where $T$ is measured in degrees Celsius; $A$ and $B$ are empirical parameters. The typical values of these parameters are about 205 W m$^{-2}$ and 2.0 W m$^{-2}$°C$^{-1}$ respectively (e.g., North and Coakley 1979; McGuffie and Henderson-Sellers 2005; Karper and Engler 2013).

We will consider the climate engineering scheme in which the climate is controlled via the injection of sulfate aerosol particles or precursor gases into the stratosphere, causing the disturbances in the radiative balance of Earth and in such a way imitating the influence of volcanic aerosols on global climate. Note that some alternatives to sulfate aerosols are also available (Jones et al. 2016; Pope et al. 2012). Injected stratospheric aerosol particles scatter shortwave solar radiation back to outer space and consequently change the radiative balance of our planet increasing the Earth's planetary albedo. Let the climate system be affected by a small radiative forcing $\delta Q$ caused by stratospheric artificial aerosols. Thus, we assume that $Q = Q_0 + \delta Q$ and $\delta Q \ll Q_0$, where $Q_0$ is the insolation corresponding to some unperturbed state of the climate system. The radiative forcing $\delta Q$ induces the small changes in planet's average surface temperature $\delta T$ such that $\delta T << T_0$, where $T_0$ is the unperturbed value of global mean surface temperature. The equation that governs perturbation $\delta T$ can be derived by linearizing equation

(1) around the unperturbed state. Applying the linearization procedure suggested by Tung (2007) and Karper and Engler (2013) we can obtain the following perturbation equation:

$$B_0 \tau \frac{d\delta T}{dt} = \left(1 - \alpha_0\right) \delta Q - \frac{B_0}{G} \delta T \ . \tag{4}$$

Here, $\tau = C/B_0$ is a relaxation time, $G$ is a climate gain such that $G = 1/\left(1 - f\right)$, where

$$f = f_1 + f_2,$$

$$f_1 = -\left(A_1 + B_1 T_0\right)/B_0,$$

$$f_2 = -\alpha_1 Q_0 / B_0 \ .$$

Climate gain factor $G$, as discussed by Tung (2007), ranges from $G \sim 1$ to 3. To derive the equation (4), we have assumed that parameters $A$ and $B$ and the albedo $\alpha$ can be expanded into a power series of the form (e.g., Karper and Engler 2013):

$$\xi: \ T \mapsto \xi\left(T\right) = \xi_0 + \xi_1 \delta T + H.O.T, \ \ \xi = A, \ B, \ \alpha \ .$$

In the model, we cannot afford to ignore the effect of radiative forcing $\Delta R_{CO_2}$ caused by the global increase in the atmospheric carbon dioxide ($CO_2$) concentration. Usually this forcing can is represented as $\Delta R_{CO_2} = \lambda \times \ln\left(C\left(t\right)/C_0\right)$, where $\lambda$ (W m$^2$) is the empirical coefficient, $C(t)$ is the $CO_2$ concentration at time $t$, and $C_0$ is the initial $CO_2$ concentration. A typical value for the parameter $\lambda$ is near 5.35 W m$^2$ (Myhre et al. 1998). However, in our model we take into account the total global mean anthropogenic and natural radiative forcing $\Delta R_{Tot}$ prescribed by the RCP scenarios. We use a linear function of time to approximate $\Delta R_{Tot}$:

$$\Delta R_{Tot}\left(t\right) = R_0 + \eta t \ , \tag{5}$$

where $R_0$ is a background radiative forcing, if there is one; and $\eta$ is the annual anthropogenic and natural radiative forcing rate estimated in accordance with the RCP scenarios (see Table 1) (e.g., Moss et al. 2008; Meinshausen et al. 2011).

Table 1. Annual natural and anthropogenic forcing $\eta$ rate obtained from the RCP scenarios (Moss et al. 2008; Meinshausen et al. 2011)

| Scenario | RCP 8.5 | RCP 6.0 | RCP 4.5 | RCP 2.6 |
|---|---|---|---|---|
| $\eta$, W m$^{-2}$ year$^{-1}$ | $7.14 \times 10^{-2}$ | $3.84 \times 10^{-2}$ | $2.17 \times 10^{-2}$ | $8.30 \times 10^{-4}$ |

As we have already mentioned, the climate control is assumed to be executed through the injection of aerosols or aerosol precursors into the stratosphere causing the radiative forcing which can be calculated via the albedo of aerosol layer:

$$\delta Q_A = -\alpha_A Q_0 \,, \qquad (6)$$

where $\delta Q_A$ is the instant radiative forcing caused by the aerosol cloud, and $\alpha_A$ is the instant albedo of this cloud. Meanwhile, $\delta Q_A$ can be also expressed through the optical depth of the aerosol layer $T_A$ (e.g., Bluth et al. 1992; Eliseev et al. 2010; Hansen et al. 2005; Lenton and Vaughan 2009):

$$\delta Q_A = -\beta_A T_A \,, \qquad (7)$$

where the coefficient $\beta_A = 24$ W m$^{-2}$ (Lenton and Vaughan 2009; Hansen et al. 2005).

In our model the albedo $\alpha_A$ is taken as a control parameter which varies in time. In fact, when executing SRM projects, we control the emission rate of the sulfate aerosols injected into the stratosphere. Suppose that when we solve the control problem, we find the time dependent albedo $\alpha_A(t)$ which is optimal. By combining (6) and (7), we can gain a simple equation that allows us to estimate the instant optical depth of the aerosol layer $T_A$ if the instant albedo $\alpha_A$ is known:

$$T_A = \left(\alpha_A / \beta_A\right) Q_0 \,. \qquad (8)$$

Respectively, the total instant mass of aerosols can be calculated as

$$M_A = \left(T_A / k_A\right) S_e \,. \qquad (9)$$

Here, $k_A$ is the mass extinction coefficient measured in m$^2$g$^{-1}$; $S_e$ is the Earth's area determined as $S_e = 4\pi R_e^2$, where $R_e = 6371$ km is the Earth's radius. From (8) and (9) we can derive the following linear relationship between the instant mass of aerosols and the instant albedo of the aerosol layer:

$$M_A = \alpha_A \left(Q_0 / \beta_A k_A\right) S_e \,. \qquad (10)$$

Thus, knowing the optimal albedo as a function of time $\alpha_A(t)$, we can determine the corresponding temporal change in the total aerosol mass $M_A(t)$. Then the aerosol emission rate $E_A(t)$ can be calculated from the following mass balance equation:

$$\frac{\partial M_A(t)}{\partial t} = E_A(t) - \frac{M_A(t)}{\tau_A} \,. \qquad (11)$$

where $\tau_A$ is the residence time of stratospheric aerosol particles. Note that a similar parameterization scheme of aerosol effects has previously been used in modelling of climate engineering (e.g., Eliseev et al. 2010).

The instant albedo $\alpha_A$ of the sulfate stratospheric aerosol layer and the corresponding calculated instant optical depth $T_A$ and instant total mass of aerosols $M_A$ are listed in Table 2. In calculations, the mass extinction coefficient $k_A$ for sulfate aerosols is assumed to be equal to 7.6 $m^2 g^{-1}$ (Eliseev et al. 2010).

Table 2. Instant albedo of stratospheric aerosol layer $\alpha_A$ and the corresponding calculated instant optical depth $T_A$ and total mass of aerosols $M_A$

| $\alpha_A$ | 0,005 | 0,010 | 0,015 | 0,020 | 0,025 |
|---|---|---|---|---|---|
| $T_A$ | 0,07 | 0,14 | 0,21 | 0,29 | 0,36 |
| $M_A$ Tg | 4,7 | 9,6 | 14,4 | 19,2 | 23,9 |

Substituting (5) and (6) into (4) we obtain the following equation to model the control system:

$$B_0 \tau \frac{d\delta T}{dt} + \frac{B_0}{G} \delta T = -\left(1-\alpha_0\right)\alpha_A Q_0 + R_0 + \eta t \; . \tag{12}$$

For the particular case $\alpha_A = const$ the analytic solution of this equation is

$$\delta T\left(t\right) = GB_0^{-1}\left[\eta\left(t-\Delta_t\right) + q\left(1-e^{-t/G\tau}\right)\right], \tag{13}$$

where $\Delta_t = G\tau\left(1-e^{-t/G\tau}\right)$ is the delay, and $q = R_0 - \left(1-\alpha_0\right)\alpha_A Q_0$ is the unchanging radiative forcing which includes the effect of stratospheric aerosols.

Effective heat capacity C can be calculated by the following formula (e.g., Fraedrich 2001):

$$C = C_A + C_O = c_p\left(p_S/g\right) + c_w \rho_w h \delta_B \; , \tag{14}$$

where $C_A$ is the effective heat capacity of the atmosphere; $C_O$, heat capacity of the ocean with mixing layer depth equal to $h$; $\delta_B$, fraction of the ocean area; $c_p$, specific heat capacity of the air at the constant pressure; $p_s$, surface pressure; $g$, gravity acceleration; $c_w$, specific heat capacity of the ocean water. For typical values of $c_p = 1004$ J $K^{-1}$ $kg^{-1}$, $p_s \approx 10^5$ Pa, $g = 9.81$ m $s^{-2}$, $c_w = 4218$ J K$^{-1}$ kg$^{-1}$, $\rho_w = 1025$ kg m$^{-3}$, $\delta_B \approx 0.71$, and $h$= 75 m (Hartmann 1994), equation (14) gives $C \approx 2.4 \times 10^8$ J m$^{-2}$ K$^{-1}$ or $C \approx 7.62$ W year m$^{-2}$ K$^{-1}$.

## 3  Optimal control without state constraint: problem statement and computation results

For the sake of convenience, let us introduce the control variable $u$ by $u \equiv \alpha_A$ and then rewrite the equation (12) as

$$\frac{d\delta T(t)}{dt} = -a\delta T(t) - bu(t) + ct + d,\tag{15}$$

where

$$a = \frac{1}{G\tau},\quad b = \frac{Q_0(1-\alpha_0)}{B_0\tau},\quad c = \frac{\eta}{B_0\tau},\quad d = \frac{R_0}{B_0\tau}.$$

By way of illustration, let us examine the so-called minimum-energy control problem that involves finding the control function, i.e. the albedo of aerosol layer, in order to minimize the following performance index:

$$J(u) \triangleq \frac{1}{2}\int_{t_0}^{t_f} u^2(t)dt\tag{16}$$

Assume the permissible control satisfies the following condition:

$$0 \le u(t) \le U \text{ for all } t \in \left[t_0,\, t_f\right],\tag{17}$$

where $U$ is the maximum value of technically feasible and affordable albedo $\alpha_A$.

The OCP ($\mathcal{P}_0$) is formulated as follows:

*Find the admissible control trajectory* $u^*: \left[t_0,\, t_f\right] \mapsto U \subset \mathbb{R}$ *engendering the corresponding state trajectory* $\delta T^*: \left[t_0,\, t_f\right] \mapsto \mathbb{R}$ *such that the performance index* $J$ *is minimized under the equality dynamic constraint* (15) *and given initial* $\delta T(t_0) = \delta T_0$ *and terminal* $\delta T(t_f) = \delta T_f$ *conditions.*

In this formulation the terminal condition $\delta T_f$ can be interpreted as a *target change in temperature* at $t = t_f$.

To be consistent with Pontryagin's maximum principle, we shall consider an equivalent maximization problem $\max_{u \in U}(-J)$ for which the Hamiltonian function is of the form:

$$H\left(\delta T, u, \psi, t\right) = -\frac{1}{2}u^2 + \psi\left(-a\delta T - bu + ct + d\right), \tag{18}$$

where $\psi \in \mathbb{R}$ is the adjoint (costate) variable. We can rewrite (18) as

$$H\left(\delta T, u, \psi, t\right) = -\frac{1}{2}\left(u - u_m\right)^2 + \frac{1}{2}b^2\psi^2 + \psi\left(-a\delta T + ct + d\right), \tag{19}$$

where $u_m(t) = -b\psi(t)$. The Hamiltonian $H$ is quadratic in $u$ and its graph is a parabola whose branches are directed downward. As shown in Fig. 1, this function attains its maximum either on the boundaries of the admissible control region (if $u_m < 0$ or $u_m > U$), or within its interior (if $0 \le u_m \le U$). Thus, the optimal control takes the form:

$$u^*(t) = \begin{cases} U, & \text{if } u_m(t) > U, \\ u_m(t), & \text{if } 0 \le u_m(t) \le U, \\ 0, & \text{if } u_m(t) < 0. \end{cases} \tag{20}$$

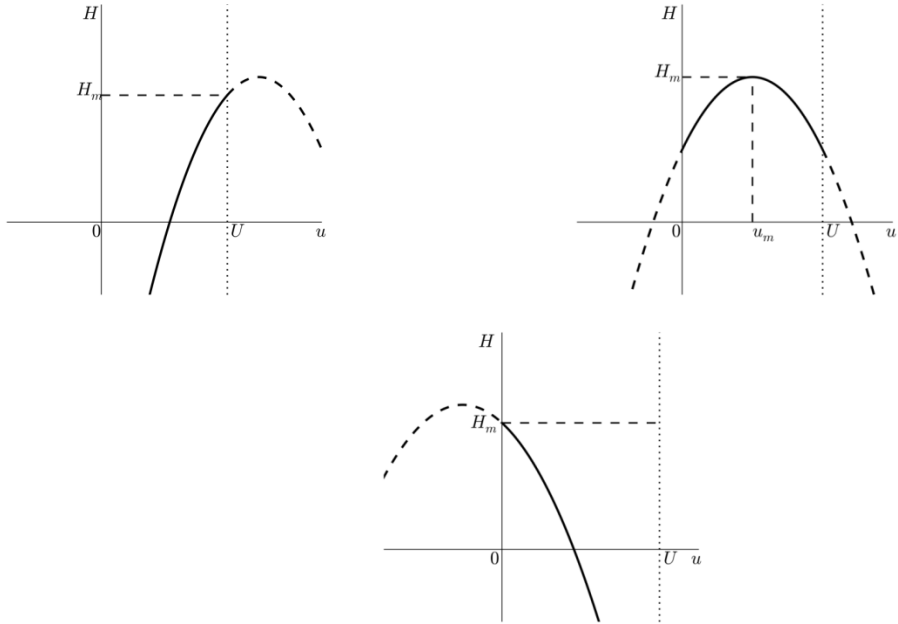

Fig. 1  Determination of maximum of the Hamiltonian function

In order to find the optimal control, we need to determine the adjoint variable $\psi(t)$. In the problem $\mathcal{P}_0$ the costate equation and the stationarity condition can be written as:

$$\frac{d\psi}{dt} = -\frac{\partial H}{\partial(\delta T)} = a\psi, \tag{21}$$

$$\frac{\partial H}{\partial u} = -u - b\psi = 0. \tag{22}$$

Since $\partial^2 H/\partial u^2 = -1$, then from (22) it follows that $H$ is maximized when

$$u(t) = -b\psi(t). \tag{23}$$

From the equation (21) we have

$$\psi(t) = C_1 e^{at}. \tag{24}$$

Substituting (23) and (24) into the equation (15), we can get its analytic solution:

$$\delta T(t) = C_1 \frac{b^2}{2a} e^{at} + C_2 e^{-at} + \frac{c}{a} t + \frac{ad - c}{a^2}. \tag{25}$$

The constants of integration, $C_1$ and $C_2$, are determined by applying the initial $\delta T_0$ and terminal $\delta T_f$ conditions. Then, using (20), (23) and (24) we can calculate the optimal control $u^*(t) = -C_1 b e^{at}$ and, finally, the optimal state trajectory $\delta T^*(t)$ by numerically solving equation (15).

Let us now discuss some results of these calculations. We take calendar years 2020 and 2100 to be $t_0$ and $t_f$ respectively. In other words, we examine the climate control problem on a finite time interval 2020 – 2100 with initial condition $\delta T_{2020} = 0$ corresponding to 2020. Accordingly, all obtained temperature changes are analysed versus $T_{2020}$.

Since the model is linear, its solution is linear as well. Fig. 2 illustrates the linear trend over time for variations in temperature $\delta T$ calculated for different RCP scenarios. The corresponding temperature changes in 2100 are shown in Table 3.

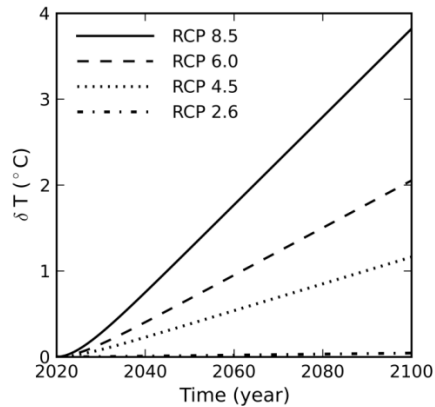

Fig. 2 Changes in global mean annual surface temperature calculated for different RCP scenarios

Table 3. Calculated temperature changes $\delta T_{2100}$ relative to $T_{2020}$

| Scenario | RCP 8.5 | RCP 6.0 | RCP 4.5 | RCP 2.6 |
|---|---|---|---|---|
| $\delta T_{2100}$, $^{o}$C | 3.82 | 2.05 | 1.16 | 0.04 |

The calculated values of $\delta T_{2100}$ agree with available estimates (e.g., Nazarenko et al. 2015). In calculations the parameter $G$ took the value of 1.5. Note that this parameter is uncertain (Tung 2007) and to some extend affects the calculated value of $\delta T_{2100}$. However, from a methodological perspective this is not very important.

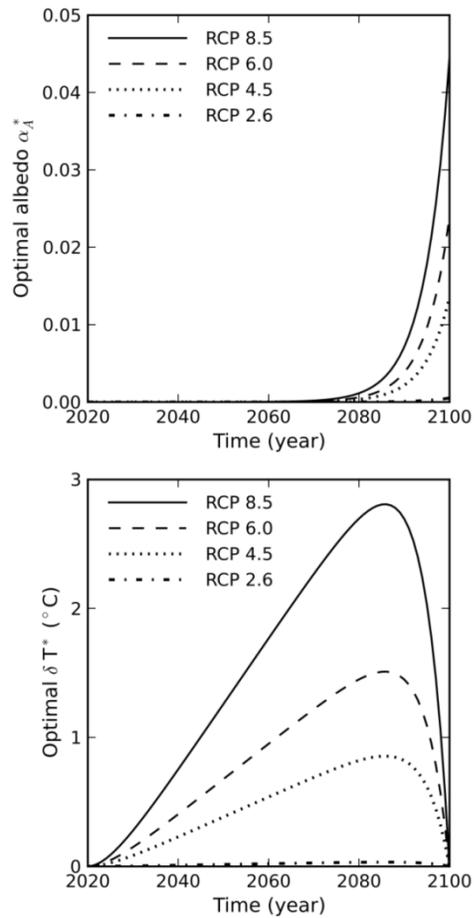

Fig. 3 Calculated optimal albedo of aerosol layer $\delta\alpha_A^*$ as a function of time (upper graph) and corresponding changes in global mean annual surface temperature $\delta T^*$ (lower graph) for different RCP scenarios and $\delta T_f = 0\,^{o}$C (without control constraint)

First, let us assume that there is no constraint on the control variable. The optimal albedo $\alpha_A^*(t)$ and the corresponding optimal temperature changes $\delta T^*(t)$ calculated for different RCP scenarios are shown in Fig. 3. These results are obtained with the terminal condition $\delta T_f = \delta T_{2100} = 0$, which means that $T_{2100} = T_{2020}$. In order to satisfy this condition, an exponential growth of the albedo $\alpha_A$ is required. Maximum values of $\alpha_A$ and, consequently, maximum values of calculated mass of aerosols $M_A$ are reached in 2100 (see Table 4).

Table 4. Values of the albedo $\alpha_{A,2100}$ and the corresponding total mass of aerosols $M_{A,2100}$ for different RCP scenarios without control constraint for terminal condition $\delta T_{2100} = 0$

| Scenario | RCP 8.5 | RCP 6.0 | RCP 4.5 | RCP 2.6 |
|---|---|---|---|---|
| $\alpha_{A,2100}$ | $4.45 \times 10^{-2}$ | $2.39 \times 10^{-2}$ | $1.35 \times 10^{-2}$ | $5.17 \times 10^{-4}$ |
| $M_{A,2100}$ Tg | 42.5 | 22.9 | 12.9 | 0.5 |

Despite the fact that the target change in temperature $\delta T_{2100} = 0$ is satisfied, there is a certain increase in temperature ("overheating") within the given time interval 2020 – 2100, i.e. $\delta T(t) > \delta T_{2100}$, where $t \in [t_0, t_f]$ (lower graph in Fig. 3). The maximum temperature increases $\delta T_{\max}$ for different RCP scenarios are presented in Table 5. In particular, for the RCP8.5 scenario the "overheating" maximum value of about 2.8℃ is reached in 2085 (see lower graph in Fig. 3). If this overheating is regarded as an undesirable phenomenon, the OCP must be considered with state constraint

$$\delta T(t) \leq C_T \quad \forall t \in [t_0, t_f], \tag{26}$$

where $C_T$ is the threshold parameter whose value should be set. The OCP with state constraint will be considered in the next Section.

Let us discuss the results obtained with control constraint (20). Suppose that $U = 0.02$. The corresponding instant total aerosol mass is estimated at 19.2 Tg (see Table 2). Fig. 4 shows the optimal albedo $\alpha_A^*(t)$ (upper graph) and the associated optimal temperature changes $\delta T^*(t)$ (lower graph) calculated for different target increases in temperature $\delta T_{2100}$ for the RCP8.5 scenario, which is a scenario of relatively high GHG emissions. From the analysis of Fig. 4 it follows that the optimal albedo $\alpha_A^*(t)$, subject to constraint $\alpha_A^*(t) \leq U$, ensures the satisfaction of the terminal condition $\delta T_{2100}$ only if $\delta T_{2100} \geq 2\,^\circ\text{C}$. If, for example, $\delta T_{2100} = 1\,^\circ\text{C}$ or $\delta T_{2100} = 0$, then the corresponding calculated optimal change in temperature in 2100 $\delta T_{2100}^*$ is about 1.4℃ and 1.2ºC respectively. In these cases, to satisfy the terminal conditions we need to increase the values of control constraint parameter $C_T$.

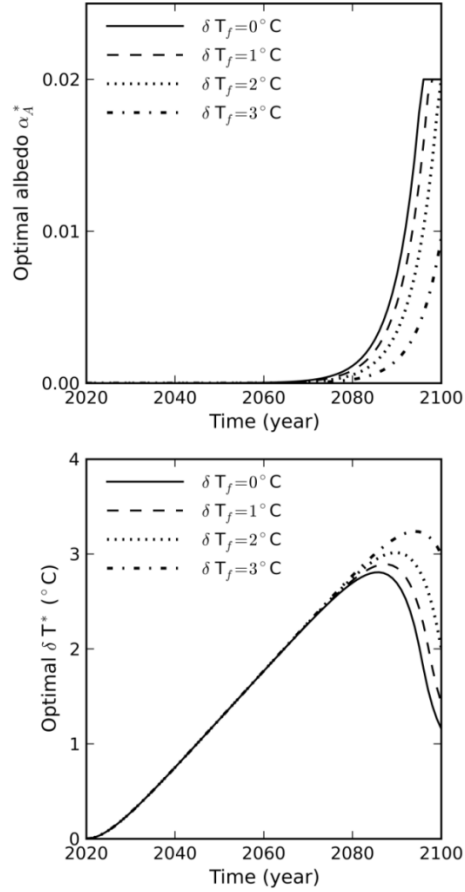

Fig. 4 Calculated optimal albedo of aerosol layer $\delta\alpha_A^*$ as a function of time (upper graph) and corresponding changes in global mean annual surface temperature $\delta T^*$ (lower graph) for different target changes in temperature $\delta T_f$ and RCP8.5 scenario with control constraint

Table 5. The maximum temperature increases $\delta T_{\max}$ for different RCP scenarios and terminal condition $\delta T_{2100} = 0$ without control constraint

| Scenario | RCP 8.5 | RCP 6.0 | RCP 4.5 | RCP 2.6 |
|---|---|---|---|---|
| $\delta T_{\max}$, ºC | 2.81 | 1.51 | 0.85 | 0.03 |

It is obvious that the value of $U$ affects the optimal state trajectory $\delta T^*(t)$. As an example, Fig. 5 shows $\delta T^*(t)$ for the RCP 6.0 scenario calculated for different values of the control constraint $U$ and for $\delta T_{2100} = 0$. Fig. 5 reveals that only sufficiently high values of the albedo of aerosol layer ($\alpha_A > 0.015$) allow the fulfilment of the terminal condition $\delta T_{2100} = 0$.

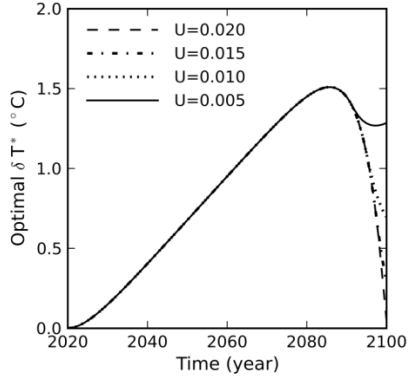

Fig. 5 Optimal changes in global mean annual surface temperature $\delta T^*$ calculated for different values of control constraint for the RCP 6.0 scenario

## 4    Optimal control with state constraint: problem formulation and calculation results

In the previous section, exploring the OCP without any constraints on state variable $\delta T(\cdot)$, we obtained that within the time interval of interest some undesirable overheating may occur, which can exceed the value of target temperature increase $\delta T_f$ (see Fig. 4 and Table 5). Assume the challenge before us is to limit this overheating by imposing the inequality constraint on state variable $\delta T(\cdot)$. In this case the OCP ($\mathcal{P}_c$) is formulated as follows:

*Minimize the objective function* (16) *subject to the inequality state constraint* (26) *and dynamical constraint* (15) *given the initial and terminal conditions*

$$\delta T\left(t_0\right)=\delta T_0,\ \delta T\left(t_f\right)=\delta T_f\ . \tag{27}$$

The constraint (26) is called pure state constraint since it does not explicitly depend on the control variable. Consequently, state variable $\delta T(\cdot)$ can be controlled only indirectly via the state equation. This creates some difficulties for solving the OCP. A usual way to overcome these difficulties is to transform the pure state constraint (26) into the corresponding mixed constraint, which explicitly depends on both the state variable and the control variable. Let us represent the inequality (26) as follows:

$$S\left(\delta T,t\right)\equiv C_T-\delta T\left(t\right)\ge 0,\ \forall t\in\left[t_0,t_f\right]. \tag{28}$$

Generally, to handle this constrain we can apply either direct or indirect methods (e.g., Bryson and Ho 1975; Sethi and Thompson 2000), which are the two classical solution techniques. Direct methods involve reducing the OCP to a nonlinear programming problem, whereas indirect methods transform the OCP to a two-point boundary value problem. However, the derivation of this problem requires a priori knowledge of the optimal solution $\left(\delta T^*, u^*\right)$ obtained without the state constraint (Bryson and Ho 1975). Since we already have this optimal solution, we can apply the indirect approach which allows us to stay within the Pontryagin's principle framework. Fig. 6 illustrates schematically two optimal state trajectories obtained correspondingly with and without the state constraint.

To proceed further, we need to make some additional explanations and informal definitions relevant to Fig. 6. With regard to the constraint (28), a certain time interval $\left(t_1, t_2\right) \subset \left[t_0, t_f\right]$, with $t_1 < t_2$, is called an interior interval if the constraint (28) is inactive, i.e. $S\left(\delta T, u\right) > 0 \ \ \forall t \in [t_1, t_2]$. Inactive constraint is not binding and, therefore, can be omitted. An interval $\left(t_1, t_2\right) \subset \left[t_0, t_f\right]$ is said to be a boundary interval if the constraint (28) is active (or tight), i.e. $S\left(\delta T, u\right) = 0 \ \ \forall t \in [t_1, t_2]$. If an interior interval ends at $t_1$ and a boundary interval begins at $t_1$ then an instant in time $t_1$ is referred to as an entry time (point). However, if there is a boundary interval ending at $t_2$, and an interior interval beginning at $t_2$, then an instant $t_2$ is said to be an exit time (point). The point in time $t_c$ is called contact time if the state trajectory $x(t)$ only touches the boundary at $t_c$, and $x(t)$ is in the interior prior to and after $t_c$. Entry, exit, and contact times are known as the junction times.

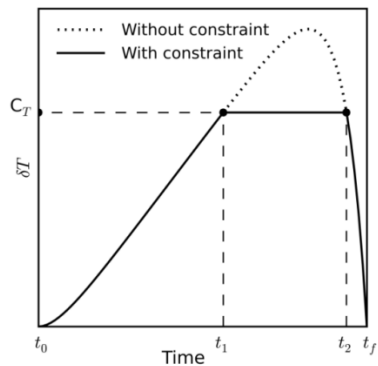

Fig. 6 Schematic illustration of optimal system trajectories with and without state constraint

To take into account whether the constraint (28) is active or inactive, we introduce the time-dependent multiplier $\mu$ such that (Bryson and Ho 1975):

$$\mu(t) \ge 0, \ \mu(t) S(\delta T, t) = 0, \tag{29}$$

where $\mu \equiv 0$ whenever $S > 0$, and $\mu \ge 0$ otherwise. Conditions (29) are really additional necessary conditions for optimality known as the *complementary slackness conditions* that must hold when solving the OCP with state constraints.

To transform the pure state constraint (28) into the corresponding mixed constraint we have to determine the order of the state constraint by differentiating $S(\delta T, t)$ with respect to time as many times as required until the constraint explicitly includes a control variable. It is fairly easy to show that the constraint (28) is of the first order since the first derivative of $S(\delta T, t)$ with respect to time has explicit dependence on $u$:

$$S^1(\delta T, u, t) = \frac{dS}{dt} = \frac{\partial S}{\partial t} + \frac{\partial S}{\partial \delta T} F(\delta T, u, t), \tag{30}$$

where $F(\delta T, u, t)$ is the right-hand side of the state equation (15).

To formulate the maximum principle the multiplier $\mu$ that directly multiplies $S^1(\delta T, u, t)$ is appended to the Hamiltonian (18), forming the augmented Hamiltonian:

$$L(\delta T, u, \psi, \mu, t) = H(\delta T, u, \psi, t) + \mu(t) S^1(\delta T, u, t), \tag{31}$$

where $S^1 = 0$ on the boundary interval, $S = 0$, and $\mu = 0$ on the interior interval, $S > 0$ (Bryson and Ho 1975). Certainly, the multiplier $\mu$ satisfies the complementary slackness conditions (29).

For the problem with state constraint the maximum principle is formulated as follows:

(i) If $u^*(t)$ is an optimal control trajectory generating the corresponding optimal state trajectory $\delta T^*(t)$ then the Hamiltonian (18) attaints its maximum with respect to $u$ for every $t \in [t_0, \ t_f]$,

$$H(\delta T^*, u^*, \psi^*, t) = \max_u H(\delta T^*, u, \psi^*, t), \tag{32}$$

(ii) Functions $\delta T^*$, $u^*$, $\psi^*$ and $\mu^*$ satisfy the Euler-Lagrange equations

$$d\delta T^* / dt = L_\psi (\delta T^*, u^*, \psi^*, \mu^*, t), \tag{33}$$

$$d\psi^* / dt = -L_{\delta T}\left(\delta T^*, u^*, \psi^*, \mu^*, t\right),\tag{34}$$

$$0 = -L_u\left(\delta T^*, u^*, \psi^*, \mu^*, t\right),\tag{35}$$

where the subscripts of $\psi$, $\delta T$ and $u$ denote corresponding partial derivatives;

(iii) The optimal multiplier $\mu$ satisfies conditions

$$\mu^* S\left(\delta T^*, t\right) = 0, \quad \mu^* \geq 0,\tag{36}$$

(iv) The state trajectory at entry and exit times satisfies the so-called tangency constraints (Bryson and Ho 1975):

$$N\left(\delta T, t\right) \triangleq \begin{bmatrix} S\left(\delta T, u\right) \\ S^1\left(\delta T, u\right) \end{bmatrix} = 0.\tag{37}$$

Let's go back to the Fig. 6. On the half-closed time interval $\left[t_0, \ t_1\right)$ the constraint (28) is not active since $S\left(\delta T, t\right) > 0$ for all $t \in \left[t_0, \ t_1\right)$. As a result, from the complementary slackness conditions it follows that $\mu(t) = 0$ if $t \in \left[t_0, \ t_1\right)$. At the entry point $t_1$ the constraint (28) is tight. Thus, on the interval $\left[t_0, \ t_1\right]$ we must solve the OCP, similar to the problem considered in the previous section, with the difference that the right boundary condition is imposed at $t = t_1$ but not at the terminal time $t_f$, i.e. the boundary conditions are:

$$\delta T\left(t_0\right) = \delta T_0, \ \delta T\left(t_1\right) = C_T.\tag{38}$$

Solving this two-point boundary value problem, we can obtain the constants of integration, $C_1$ and $C_2$, the entry time, $t_1$, and, consequently, the optimal control $u^*\left(t\right)$ and optimal state trajectory $\delta T^*\left(t\right)$ for all $t \in \left[t_0, \ t_1\right]$. Similarly, we solve the OCP on the time interval $t \in \left[t_2, \ t_f\right]$, where $t_2$ is the exit point. In this case a two-point boundary value problem is solved with the following left and right boundary conditions:

$$\delta T\left(t_2\right) = C_T, \ \delta T\left(t_f\right) = \delta T_f.\tag{39}$$

If the state trajectory reaches the entry point $t_1$, where $S(\delta T, t_1) = 0$, then for the orbit to remain on the boundary over the time interval $t \in \left(t_1, \ t_2\right)$, $S^1(\delta T, t) = 0$ is required for all $t \in \left(t_1, \ t_2\right)$. Thus, the optimal control can be found from the following equation:

$$0 = -aC_T - bu^*(t) + ct + d, \ \forall t \in \left(t_1, \ t_2\right).\tag{40}$$

Figs. 7 and 8 illustrate the simulation results for RCP8.5 and RCP6.0 scenarios respectively obtained for two constraint parameter values: $C_T = 0.5\ ^oC$ (upper graphs) and $C_T = 1\ ^oC$ (lower graphs). The optimal control trajectories $\alpha_A^*(t)$ as well as the corresponding optimal state trajectories $\delta T^*(t)$ are shown in these figures.

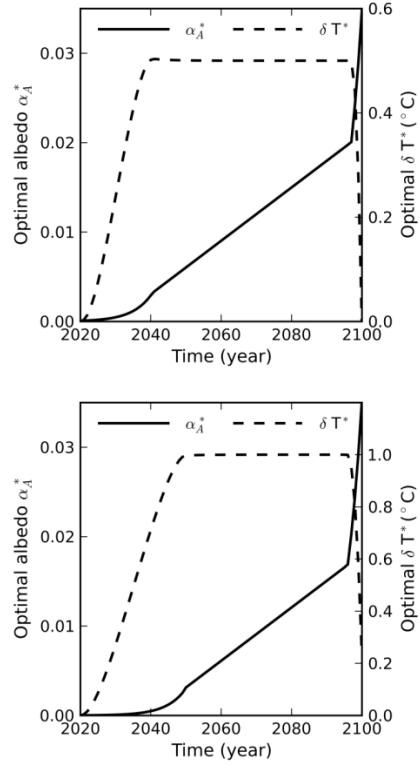

Fig. 7 Calculated optimal albedo of aerosol layer $\delta\alpha_A^*$ as a function of time and corresponding changes in global mean annual surface temperature $\delta T^*$ for RCP 8.5 scenario for two state constraint values: $C_T = 0.5\ ^oC$ (upper graph); $C_T = 1.0\ ^oC$ (lower graph)

As mentioned above, in the model the albedo $\alpha_A$ is considered to be a control variable, however, in actuality we control the aerosol emission rate $E_A$. Using equation (11) we can estimate $E_A$ that corresponds to the optimal albedo assuming the residence time of stratospheric aerosol particles $\tau_A$ is two years (Hansen et al. 1992), and taking into account that one tonne of sulfur released annually into the stratosphere forms approximately four tonnes of aerosol particles (Rasch et al. 2008). Therefore, instead of the emission rate of aerosol particles $E_A$, we can consider the emission rate of sulfur $E_S$. Let $\bar{E}_S$ denote the annual emission rate of sulfur averaged over the period 2020 – 2100. For the RCP8.5 scenario and $C_T = 1^\circ\text{C}$ (lower graph in Fig. 7) we obtained $\overline{E}_S \approx 1$ Tg year$^{-1}$. The maximum value $\overline{E}_{S,\max} \approx 5.6$ Tg year$^{-1}$ is reached in 2100. The total mass of sulfur $M_S$ loaded during the specified time period is about 80 Tg. For the RCP6.0 scenario and $C_T = 1^\circ\text{C}$ (lower graph in Fig. 8), we obtained $\overline{E}_S \approx 0.36$ Tg year$^{-1}$ with the maximum value $\overline{E}_{S,\max} \approx 3.4$ Tg year$^{-1}$ attained in 2100, and $M_S \approx 30$ Tg.

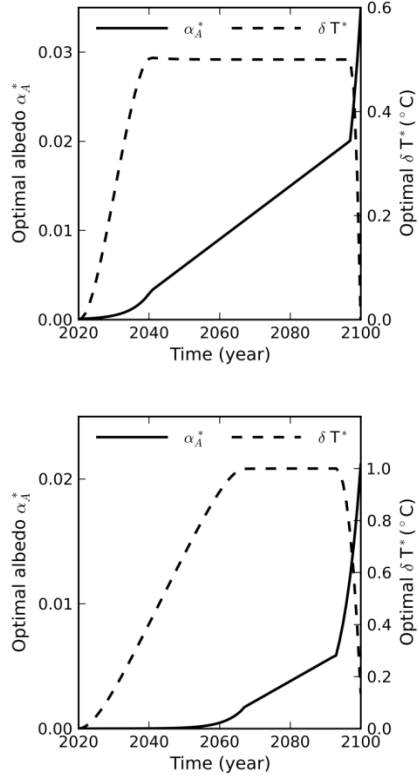

Fig. 8 Calculated optimal albedo of aerosol layer $\delta\alpha_A^*$ as a function of time and corresponding changes in global mean annual surface temperature $\delta T^*$ for RCP 6.0 scenario for two state constraint values: $C_T = 0.5\,^\circ\text{C}$ (upper graph); $C_T = 1.0\,^\circ\text{C}$ (lower graph)

It is clear that the calculated emission rates and the mass of sulfur depend on the extinction coefficient $k_A$ and the parameter $\beta_A$ (see equation (10) in Section 2). Nevertheless, using the values of these parameters, given in Section 2, the estimates wecalculated are in satisfactory agreement with the estimates obtained earlier by other researchers (e.g., Crutzen 2006; Brovkin et al. 2009; Eliseev et al. 2010). However, unlike previous explorations of climate engineering, in this paper the optimal emission scenario is a solution to the optimal control problem, rather than known a priori.

# 5    Concluding remarks

Scientific community considers the use of fine aerosol particles, artificially injected into the stratosphere, to be one of the most effective and feasible measures to counter global warming. Computer simulation using mathematical climate models of various degrees of sophistication and complexity is the most popular and reliable technique for exploring and estimating the effectiveness of stratospheric aerosol climate engineering. Numerical simulation of climate engineering requires the development of somewhat realistic scenarios for aerosol injection. Undoubtedly, the development of scenarios is a subjective process, since each researcher has the right to choose those scenarios that are the most realistic from his point of view. However, the subjectivity of scenario formulation to some extent reduces the value of the results obtained. Besides, while exploring the geoengineering methods, we make a major assumption that mankind is actually able to some extent control the behavior of the climate system in our pursuit to achieve desired results. Unfortunately, the objectives of geoengineering projects are usually formulated verbally, and the problem of climate engineering is considered outside the framework of control theory.

This paper demonstrates the application of the theory of optimal control to climate engineering problems. In this context, the objective of climate engineering is formulated not verbally, but in terms of an extremal problem, and the optimal scenario for aerosol injection is obtained as a solution to the problem of optimal control. The simplest model of the climate system, namely the zero-dimensional energy balance model, is used as a model of control system. This allows us to analytically derive the optimal control trajectory, i.e. changes over time in the albedo of the aerosol layer. Nevertheless, it is necessary to apply numerical methods to calculate the corresponding optimal state trajectory. Optimal solutions to the problem of climate engineering, both in the absence and in the presence of constraints on the control and phase variables, are obtained on the basis of the classical Pontryagin's maximum principle. The obtained aerosol emission estimates do not contradict and in fact generally agree with the presently available estimates.

In our opinion, the use of methods of the optimal control theory can provide additional helpful information for the development of optimal strategies for climate manipulation in order to counter global warming. It is important to emphasize that, unlike previous studies of SRM projects in which hypothetical aerosol emission scenarios are given a priori, in this paper we shown how the optimal emission scenario can be obtained by solving an optimal control problem with state and control constraints.

# References

Bellman R (1957) Dynamical programming. Princeton University Press, Princeton NJ, 392

Bluth GJS, Doiron SD, Schnetzler CC, Krueger AJ, Walter L S (1992) Global tracking of the SO2 clouds from the June, 1991 Mount-Pinatubo eruptions. Geophys Res Lett 19: 151-154

Brock W, Enström G, Grass D, Xepapadeas A (2013) Energy balance climate models and general equilibrium optimal mitigation policies. Journal of Economic Dynamics and Control37: 2371-2396

Brovkin V, Petoukhov V, Claussen M, Bauer E, Archer D, Jaeger C (2009) Geoengineering climate by stratospheric sulfur injections: Earth system vulnerability to technological failure. Climatic Change 92:243-259

Bryson AE, Ho Y-C (1975) Applied optimal control: Optimization, estimation and control. CRC Press, New York, 482

Bubnicki Z (2005) Modern control theory. Springer-Verlag, Berlin, 421

Budyko MI (1974) Climate and life. Academic Press, New York, 508

Coddington EA, Norman L (1955). Theory of differential equations.McGraw-Hill,New York, 429

Crutzen P (2006) Albedo enhancement by stratospheric sulfur injections: A contribution to resolve a policy dilemma? Climatic Change 77: 211–220

Dijkstra HA (2013) Nonlinear climate dynamics. Cambridge University Press, Cambridge, 367

Eliseev AV, Chernokulsky AV, Karpenko AA, Mokhov II (2010) Global warming mitigation by sulfur loading in the stratosphere: dependence of required emissions on allowable residual warming rate. Theoretical and Applied Climatology 101: 67–81

Fraedrich K (2001) Simple climate models. In: Imkeller P, von Storch JS (eds.). Stochastic climate models. Birkhäuser, Basel, pp 65 – 100

Gaskarov DV, Kisselev VB, Soldatenko SA, Stroganov VI, Yusupov RM (1998) An introduction to geophysical cybernetics and environmental monitoring. St. Petersburg State University Press, St. Petersburg, 165 (in Russ.)

Goodell J (2011)  How to cool the planet: Geoengineering and the audacious quest to fix Earth's climate. Mariner Books, Boston, 276

Greiner A, Gruene L, Semmler W (2014) Economic growth and the transition from non-renewable to renewable energy. Environment and Development Economics 19: 417-439

Hansen J, Lacis A, Rind D, Russell G, Stone P et al (1984) Climate sensitivity: analysis of feedback mechanisms. In: Hansen JE, TakahashiT (eds) Climate processes and climate sensitivity. American Geophysical Union, Washington DC, pp 130-163

Hansen J, Lacis A, Ruedy R, Sato M (1992) Potential climate impact of Mount Pinatubo eruption. Geophys Res Lett 19: 215-218

Hansen J, Sato M, Ruedy R et al (2005) Efficacy of climate forcing. J Geophys Res 110: D18104

Hartmann DL (1994) Global physical climatology. Academic Press, New York, 408

IPCC, 2013: Climate Change 2013: The Physical Science Basis. Contribution of Working Group I to the Fifth Assessment Report of the Intergovernmental Panel on Climate Change (2013). Stocker TF, Qin D, Plattner G-K, Tignor M, Allen SK, Boschung J, Nauels A, Xia Y, Bex V,Midgley PM (eds). Cambridge University Press, Cambridge, UK and New York, NY, USA, 1535

Isidori A (1999) Nonlinear control systems. Springer-Verlag, London, 293

Jarvis AJ, Leedal DT, Taylor CJ, Young PC (2009) Stabilizing global mean surface temperature: a feedback control perspective. Environmental Modelling Software, 24: 665-674

Jarvis AJ, Young PC, Leedal DT, Chotai A (2008) A robust sequential $CO_2$ emissions strategy based on optimal control of atmospheric $CO_2$ concentrations. Climatic Change 86: 357-373

Jarvis A, Leedal D (2012) The geoengineering model intercomparison project (GeoMIP): a control perspective. Atmos Sci Lett 13: 157-163

Jones AC, Haywood JM, Jones A (2016) Climatic impacts of stratospheric geoengineering with sulfate, black carbon and titania injection. Atmos Chem Phys 16:2843–2862

Karper H, Engler H (2013) Mathematics and climate. SIAM, Philadelphia, 295

Keith D (2013) A case for climate engineering (Boston Review Books). MIT Press, Boston, 224

Kirk D (1970) Optimal control theory: An introduction, Prentice Hall, Englewood Cliffs, NJ, 452

Kravitz B, Robock A, Tilmes S, Boucher O, English JM, Irvine PJ, Jones A, Lawrence MG, MacCracken M, Muri H, Moore JC, Niemier U, Phipps SJ, Sillmann J, Storelvmo T, Wang H, Watanabe S (2015) The Geoengineering Model Intercomparison Project Phase 6 (GeoMIP6): simulation design and preliminary results. Geoscientific Model Development 8: 3379-3392

Launder B, Thompson JMT (2010) Geo-engineering climate change: Environmental necessity or pandora's box? Cambridge University Press, Cambridge, 332

Lenton TM, Vaughan NE (2009) The radiative forcing potential of different climate geoengineering options. Atmos Chem Phys 9: 5539-5561

MacCracken MC (2009) On the possible use of geoengineering to moderate specific climate change impacts. Env Res Lett 4: 1-14

MacMartin DG, Kravitz B, Keith DW, Jarvis A (2014) Dynamics of the coupled human-climate system resulting from closed-loop control of solar geoengineering. Clim Dyn 43: 243-258

Maurer H, Preuss JJ, Semmler W (2013) Optimal control of growth and climate change - Exploration of scenarios. In: Cuaresma JC, Palokangas T, Tarasyev A (eds) Green growth and sustainable development. Springer-Verlag, Berlin, pp 113-139

McClellan J, Keith DW, Apt J (2012) Cost analysis of stratospheric albedo modification delivery systems. Env Res Lett 7: 034019

McGuffie K, Henderson-Sellers A (2005) A climate modelling primer, 3d ed. Wiley, New York, 287

Meinshausen M, Smith SJ, Calvin K et al (2011) The RCP greenhouse gas concentrations and their extensions from 1765 to 2300. Climatic Change 109: 213-241

Ming T, de Richter R, Liu W, Caillol S (2014) Fighting global warming by climate engineering: Is the Earth radiation management and the solar radiation management any option for fighting climate change? Renewable and Sustainable Energy Reviews 31: 792-834

Moles CG, Lieber AS, Banga JR, Keller K (2003) Global optimization of climate control problems using evolutionary and stochastic algorithms. In: Benitez JM, Cordon O, Hoffmann F, Roy R (eds) Advances in soft computing: Engineering design and manufacturing. Springer-Verlag, London, 331-342

Moss R, Babiker M, Brinkman S et al. (2008) Towards new scenarios for analysis of emissions, climate change, impacts, and response strategies. Intergovernmental Panel on Climate Change, Geneva, 132 pp.

Myhre G, Highwood EJ, Shine KP, Stordal F (1998) New estimates of radiative forcing due to well mixed greenhouse gases. Geophys Res Lett 25: 2715-2718

Nazarenko L, Schmidt GA, Miller RL (2015) Future climate change under RCP emission scenarios with GISS ModelE2. J Adv Model Earth Syst 7:244–267

North GR, Coakley JA (1979) Differences between seasonal and mean annual energy balance model circulations of climate and climate change. J Atmos Sci 36: 1189-1204

Peng G, LeslieL M, Shao Y (2002) Environmental modelling and prediction. Springer-Verlag, Berlin-Haidelberg, 480

Pope FG, Braesicke P, Grainger RG, Kalberer M, Watson IM, Davidson PJ, Cox RA (2012) Stratospheric aerosol particles and solar-radiation management. Nature Climate Change 2: 713-719

Pontryagin LS, Bolryanskii VG, Gamktelidze RV, Mishchenko EF (1962) The mathematical theory of optimal processes. Wiley, New York, 360

Rasch PJ, Tilmes S, Turco R, Robock A, Oman L, Chen C-C, Stenchikov GL, Garcia RR (2008) An overview of geoengineering of climate using stratospheric sulphate aerosols. Phil Trans R Soc. A 366: 4007-4037

Robock A, Marquardt A, Kravitz B, Stenchikov G (2009) Benefits, risks, and costs of stratospheric geoengineering. Geophys Res Lett 36: D19703

Roe G (2009) Feedbacks, timescales and seeing red. Ann Rev Earth and Planetary Sci 37:93–115

Schellnhuber H-J, Kropp J (1998) Geocybernetics: controlling a complex dynamical system under uncertainty. Naturwissenschaften Review Articles 85: 411-425

Shepherd J, Caldeira K, Haigh J, Keith D, Launder B, Mace G, MacKerron G, Pyle J, Rayner S, Redgwell C, Cox P, Watson A (2009) Geoengineering the climate: Science, governance and uncertainty. The Royal Society, 98

Sethi SP, Thompson GL (2000) Optimal control theory: Application to management science and economics. Springer-Verlag, New York, 506

Soden BJ, Held IM (2006) An assessment of climate feedbacks in coupled ocean-atmosphere models. J Clim 19:3354–3360

Soldatenko S, Yusupov R (2015a) An optimal control problem formulation for the atmospheric large-scale wave dynamics. Appl Math Sci 9: 875-884

Soldatenko S, Yusupov R (2015b) On the possible use of geophysical cybernetics in climate manipulation (geoengineering) and weather modification. WSEAS Transactions in Environment and Development 11: 116-125

Soldatenko S, Yusupov R (2016) The determination of feasible control variables for geoengineering and weather modification based on the theory of sensitivity in dynamical systems. Journal of Control Science in Engineering 2016: 1547462

Soldatenko SA (2017) Weather and climate manipulation as an optimal control for adaptive dynamical systems. Complexity. 2017: 4615072

Sontag ED (1990) Mathematical control theory: deterministic finite dimensional systems. Springer-Verlag, New York, 531

Stilgoe J (2015) Experiment Earth: Responsible innovation in geoengineering. Routledge, London and New York, 258

Tol RSJ (2002) Welfare specifications and optimal controlof climate change: an application of fund. Energy Economics, 24: 367-376

Tung KK (2007) Simple climate modeling. Discrete and Continuous Dynamical Systems. Series B 7: 651-660

Wigley TM (2006) A combined mitigation/geoengineering approach to climate stabilization. Science 314: 452–455

Yusupov RM (1979) Theoretical bases of control of geophysical processes. Ministry of Higher Education Publ, Moscow, 874 (in Russ.).

Zabczyk J (2007) Mathematical control theory. Birkhäuser, Boston, 260

# Appendix

The statement of OCP for any kinds of systems (technical, natural, socio-economic, etc.) or processes (deterministic or stochastic) necessitates, first of all, developing the mathematical model of a system or process to be controlled, specifying the control objectives, and imposing the physical and some others constraints that must be satisfied. It is well known that the earth's climate is a complex nonlinear dynamical system (e.g., Dijkstra 2013; Karperand & Engler 2013) that can be modelled with nonlinear partial differential equations (PDEs). To generically formulate the OCP, for convenience's sake, these PDEs, using some technique (e.g., Galerkin procedure), can be converted into a set of ordinary differential equations (ODEs), which, in certain sense, are equipollent to the original PDEs.

Let usconsider an abstract deterministic dynamical system that evolves over some time interval $[t_0, t_f]$, where the final (or terminal) time $t_f > 0$ can either be free or specified. Without loss of generality, we can consider only the case where $t_f$ is fixed. The state of a system at any instant of time $t \in [t_0, t_f]$ can be characterized by the vector $\mathbf{x} = (x_1, \ldots, x_n)^{\mathrm{T}} \in \mathbb{R}^n$, where T denotes transpose. We shall suppose that the system is controllable, i.e. the system can be moved from some initial state to any other (desired) state in a finite time using certain external manipulations characterized formally by the vector of control variables $\mathbf{u} = (u_1, \ldots, u_m)^{\mathrm{T}} \in \mathbb{R}^m$. By a system we shall understand the set of ODEs of the form:

$$\dot{x}(t) = f\big(x(t), u(t), t\big), \tag{A1}$$

where the overhead dot denotes differentiation with respect to time; $f : [t_0, t_f] \times \mathbb{R}^n \times \mathbb{R}^m \to \mathbb{R}^n$ is a given vector-valued function with components $f_i$ $(i = 1, 2, \ldots, n)$ satisfying conditions that ensure the existence and uniqueness of the solution of ODEs (1) for specified initial conditions (e.g., Coddington and Norman 1995)

$$x(t_0) = x_0. \tag{A2}$$

Generally, various physical constraints expressed as equalities and / or inequalities can be imposed on state variables. This formally means that the state vector is required to belong to a certain phase-space domain:

$$x : [t_0, t_f] \mapsto x(t) \in X \tag{A3}$$

where $X$ is a given subset of $\mathbb{R}^n$.

Control variables can also be restricted to some control domain $U$ known as a set of admissible controls such that

$$u : [t_0, t_f] \mapsto u(t) \in U \subset \mathbb{R}^m. \tag{A4}$$

In the majority of physical applications it is sufficient to consider the class of piecewise continuous and bounded controls.

Equations (A1), (A3) and (A4) when considered together limit the set of admissible terminal values of state variables. Formally, this can be indicated as:

$$x\left(t_f\right) \in X_f\left(t_f\right),$$ (A5)

where $X_f\left(t_f\right)$ is the so-called reachable set of the state variables. In other words, $X_f\left(t_f\right)$ is a set of admissible terminal values of the state variables, which can be attained when the state vector $x(\cdot)$ obeys (A1) and (A3), and the control vector $u(\cdot)$ obeys (A4). For the simplicity of presentation we consider the OCP without state and control constraints. To compare the admissible controls to each other and to determine the best control strategy we shall introduce an objective function (performance index):

$$J\left(x(\cdot), u(\cdot)\right) \triangleq \varphi\left(x_f\right) + \int_{t_0}^{t_f} \ell\left(x(t), u(t), t\right) dt,$$ (A6)

where $\varphi : \mathbb{R}^n \to \mathbb{R}$ is the terminal or endpoint cost, with $x_f = x\left(t_f\right)$ the state vector at terminal time $t_f$. The second term in equation (A6) $\ell : \mathbb{R}^n \times U \times \mathbb{R} \to \mathbb{R}$ is the running cost.

The function $J$ in the form (A6) is a classical performance index used widely in mathematical optimization and control theory, and corresponds to the so-called Boltza problem. Two special cases of this problem, the problem of *Lagrange* ($\varphi \equiv 0$) and the problem of *Mayer* ($\ell \equiv 0$), can also be of interest in the climate control theory. We need to emphasize that the formulation of objective function depends generally on the problem under consideration.

The OCP is defined as follows:

*Find the control vector trajectory* $u^* : [t_0, t_f] \mapsto U$ *engendering the corresponding state orbit* $x^* : [t_0, t_f] \mapsto R^n$ *, determined by the state equation* (A1)*, such that the objective function* (A6) *attains its extremum and the final state constraint* (A5) *is satisfied.*

Optimal control problems, depending on their complexity, can be solved numerically or analytically by the use of fundamental mathematical methods such as the classical variational calculus, the Pontryagin's maximum principle, and the Bellman's principle of optimality known as dynamic programming. In this paper, we apply the Pontryagin's principle, which gives the basic necessary conditions for optimality.

To be specific, let us suppose that we seek to find the control $u^*$ that provides the maximum of the objective function (A6), i.e. $J(u^*) = \max\limits_{u \in U} J(u)$. According to Pontryagin's maximum principle, we shall introduce the Hamiltonian function $H$ corresponding to the system (A1) and defined for all $t \in [t_0, \ t_f]$ by:

$$H(x(t), u(t), \psi(t), t) = \ell(x(t), u(t), t) + \psi^{\mathrm{T}}(t) f(x(t), u(t), t),\tag{A7}$$

where $\psi \in \mathbb{R}^n$ is the vector of time-varying Lagrange multipliers known as costate (or adjoint) vector of a system.

Pontryagin's maximum principle states that if $u^*(t)$ is an optimal control trajectory and $x^*(t)$ is the corresponding optimal state trajectory, determined by the equation

$$x^*(t) = f(x^*(t), u^*(t), t),\tag{A8}$$

then the Hamiltonian is globally maximized with respect to $u(\cdot)$ almost everywhere in the interval $[t_0, \ t_f]$ for all admissible controls $u \in U$ (Pontryagin et al. 1962):

$$H(x^*(t), u^*(t), \psi^*(t), t) = \max\limits_{u \in U} H(x^*(t), u, \psi^*(t), t),\tag{A9}$$

where $\psi^*(t)$ is the optimal costate trajectory. Thus, we have

$$u^*(t) = \arg\max\limits_{u \in U} H(x^*(t), u, \psi^*(t), t)\tag{A10}$$

The set of first-order necessary conditions for optimality are obtained by maximizing $H$ with respect to $u(\cdot)$ at $u^*(\cdot)$, which gives:

$$\dot{x} = f(x, u, t) = \left(\frac{\partial H}{\partial \psi}\right)^{\mathrm{T}} \quad \text{with } x(t_0) = x_0\tag{A11}$$

$$\dot{\psi} = -\left(\frac{\partial H}{\partial x}\right)^{\mathrm{T}} \quad \text{with } \psi(t_f) = \left.\frac{\partial \varphi}{\partial x}\right|_{x_f}\tag{A12}$$

$$\left(\frac{\partial H}{\partial u}\right)^{\mathrm{T}} = 0\tag{A13}$$

These conditions known as Euler-Lagrange equations define a two-point boundary value problem. For the stationary point (A13) to be a local maximum, it is required that the following generalized Legendre-Clebsch condition

$$H_{uu} \equiv \nabla_u^2 H \leq 0\tag{A14}$$

must hold for each t $\in \left[ t_0, t_f \right]$, where $H_{uu}$ is the Hessian or curvature matrix. If $H_{uu}$ is positive definite, the critical point is a local minimum. The condition (A14) is known as the second order condition for optimality.

We should make several important remarks.

1. Under the appropriate concavity conditions the necessary conditions of optimality are also sufficient for a global optimum.

2. The maximization problem $J \to \max\limits_{u \in U}$ can be easily transformed to an equivalent minimization one $J_1 \to \min\limits_{u \in U}$, where $J_1 = -J$.

3. To solve optimal control problems subject to control and state constraints we need to formulate additional necessary conditions (e.g., the complementary slackness conditions).